\documentclass[pra,twocolumn,showpacs,superscriptaddress,10pt]{revtex4-1}
\usepackage[colorlinks,citecolor=blue,linkcolor=red,dvipdfm]{hyperref}
\usepackage{amsmath,amssymb,graphicx}
\usepackage{times}

\begin{document}

\title{Flat-floor bubbles, dark solitons, and vortices stabilized by inhomogeneous nonlinear media}

\author{Liangwei Zeng}
\affiliation{Key Laboratory of Optoelectronic Devices and Systems of Ministry of Education and Guangdong Province,
Shenzhen Key Laboratory of Micro-Nano Photonic Information Technology,
College of Physics and Optoelectronic Engineering, Shenzhen University, Shenzhen 518060, China}

\author{Boris A. Malomed}
\email{malomed@tauex.tau.ac.il}
\affiliation{Department of Physical Electronics, School of Electrical Engineering,
Faculty of Engineering, and the Center for Light-Matter Interaction, Tel Aviv University, P.O.B. 39040, Ramat Aviv, Tel Aviv, Israel}
\affiliation{Instituto de Alta Investigaci\'{o}n, Universidad de Tarapac\'{a}, Casilla 7D, Arica, Chile}

\author{Dumitru Mihalache}
\affiliation{Horia Hulubei National Institute of Physics and Nuclear Engineering, Magurele, Bucharest, RO-077125, Romania}

\author{Yi Cai}
\affiliation{Key Laboratory of Optoelectronic Devices and Systems of Ministry of Education and Guangdong Province,
Shenzhen Key Laboratory of Micro-Nano Photonic Information Technology,
College of Physics and Optoelectronic Engineering, Shenzhen University, Shenzhen 518060, China}

\author{Xiaowei Lu}
\affiliation{Key Laboratory of Optoelectronic Devices and Systems of Ministry of Education and Guangdong Province,
Shenzhen Key Laboratory of Micro-Nano Photonic Information Technology,
College of Physics and Optoelectronic Engineering, Shenzhen University, Shenzhen 518060, China}

\author{Qifan Zhu}
\affiliation{Key Laboratory of Optoelectronic Devices and Systems of Ministry of Education and Guangdong Province,
Shenzhen Key Laboratory of Micro-Nano Photonic Information Technology,
College of Physics and Optoelectronic Engineering, Shenzhen University, Shenzhen 518060, China}

\author{Jingzhen Li}
\email{lijz@szu.edu.cn}
\affiliation{Key Laboratory of Optoelectronic Devices and Systems of Ministry of Education and Guangdong Province,
Shenzhen Key Laboratory of Micro-Nano Photonic Information Technology,
College of Physics and Optoelectronic Engineering, Shenzhen University, Shenzhen 518060, China}

\begin{abstract}
We consider one- and two-dimensional (1D and 2D) optical or matter-wave
media with a maximum of the local self-repulsion strength at the center, and
a minimum at periphery. If the central area is broad enough, it supports
ground states in the form of flat-floor \textquotedblleft bubbles", and
topological excitations, in the form of dark solitons in 1D and vortices
with winding number $m$ in 2D. Unlike bright solitons, delocalized bubbles
and dark modes were not previously considered in this setting. The ground
and excited states are accurately approximated by the Thomas-Fermi
expressions. The 1D and 2D bubbles, as well as vortices with $m=1$, are
completely stable, while the dark solitons and vortices with $m=2$ have
nontrivial stability boundaries in their existence areas. Unstable dark
solitons are expelled to the periphery, while unstable double vortices split
in rotating pairs of unitary ones. Displaced stable vortices precess around
the central point.
\end{abstract}

\maketitle

\section{Introduction}

The consideration of models based on the nonlinear Schr\"{o}dinger (NLS) equations with a spatial variation of the local coefficient in front of the cubic term makes it possible to essentially expand the variety of stable localized and extended states supported by the NLS equations. In many cases, the variation is represented by \textit{nonlinear lattices}, i.e., spatially periodic modulations of the local nonlinearity strength \cite{review}. These arrangements help to stabilize multidimensional and spatiotemporal self-trapped states in optics \cite{NLSE0,NLSEND}, photonics \cite{MVRS,NLSE3}, and various setups for Bose-Einstein condensates (BECs) in ultracold atomic gases \cite{RRP2017,soliton-NRP,RJP2019}. The interplay of
such structures with parity-time symmetry \cite{PTREV} and gauge fields \cite{NLSE1,NLSE2} was considered too. In this connection, it is relevant to mention that linear lattices, i.e., spatially periodic linear potentials, are commonly used for the creation and stabilization of solitons. In particular, linear lattices in combination with self-repulsive cubic nonlinearity \cite{Morsch} give rise to various families of gap solitons, including fundamental \cite{GAP0,Kiv1}, subwavelength \cite{GAP1,GAP2}, parity-time-symmetric \cite{GAP3,GAPRRP}, subfundamental \cite{subfund}, composite \cite{GAP7,GAP5}, surface \cite{GAP8}, moving \cite{HS1,GAP9}, dark \cite{GAP10}, quasi-discrete \cite{additional,GAP11}, and multipole \cite{GAP12} ones, as well as clusters built of them \cite{GAP13}. Further, gap solitons were predicted in moir\'{e} lattices \cite{GAP6,GAPNP} and in systems with quintic and cubic-quintic nonlinearities \cite{GAP4}. Two-dimensional (2D) gap solitons \cite{Kiv2} and vortex solitons of the gap type \cite{HS2} were predicted too.

Nonlinear lattices \cite{review} can also support many types of soliton families, including those of fundamental, dipole, and multipole types in regular \cite{NL1,Abdullaev0,Kevrekidis}, random \cite{Abdullaev1} and defect-carrying \cite{DENL} lattices, as well as in combined linear-nonlinear ones \cite{anti}, and in nonlinear lattices embedded in a space of fractional dimension \cite{NL2,CQNL}. Vortex solitons in nonlinear lattices were predicted too \cite{NL3}.

Parallel to that, experiments with bound states of solitons (alias \textquotedblleft soliton molecules") in fiber lasers \cite{EXP3} and microresonators \cite{EXP4} have drawn much interest. The experiments have explicitly demonstrated various dynamical effects, such as the buildup of the \textquotedblleft molecules" \cite{EXP2}, excitation of internal modes in them \cite{EXP1}, and switching of bound states \cite{EXP5}.

NLS equations with spatially modulated self-defocusing nonlinearity constitute another class of models that support various localized and extended states \cite{review}. These include strongly \cite{SDN1,SDN3} and weakly \cite{SDN2} localized fundamental and multipole solitons, self-trapped states in photonic-crystal fibers \cite{SDN4}, dark solitons with slowly decaying tails \cite{Jinhua}, solitary vortices created by the application of a torque \cite{SDN6}, \textit{hybrid vortices} \cite{SDN5}, localized states in a spin-orbit-coupled system \cite{Foshan}, flat-top modes \cite{SDN7,SDN8}, \textit{hopfions} \cite{SDN9}, soliton gyroscopes \cite{SDN10}, non-autonomous solitons \cite{SDN11}, and vortex clusters \cite{SDN12,SDN13}. A cardinal difference from the settings based on focusing nonlinearities is that self-trapping of the various modes may be provided, counter-intuitively, by the spatial modulation of local self-repulsion. The use of the local nonlinearity with the defocusing sign makes such models quite promising for the creation of \emph{stable} 2D and 3D bright solitons. Although similar states can be readily found as stationary solutions in the case of the self-focusing nonlinearity, they are vulnerable to the collapse-induced instability \cite{COLLAPSE}. Obviously, the collapse does not occur in the case of the self-defocusing. Furthermore, the use of the spatially modulated self-repulsion opens the way to create complex 3D dynamical states which, otherwise, cannot exist, such as the above-mentioned single-component hopfions \cite{SDN9} and hybrid vortices \cite{SDN5}.

\begin{figure*}[tbp]
\begin{center}
\includegraphics[width=1.6\columnwidth]{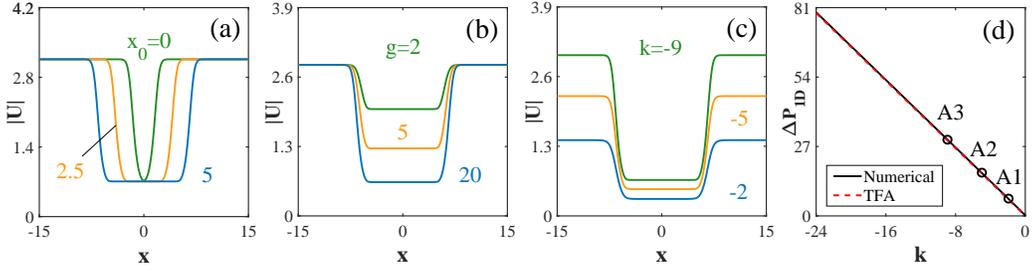}
\end{center}
\caption{(a) The transition of ordinary (relatively narrow) 1D \textquotedblleft bubbles" into flat-floor ones with the increase of the width of the nonlinearity profile in the 1D version of Eq. (\protect\ref{gprofile}), i.e., for values $x_{0}=0,2.5,5$, at a fixed nonlinearity strength, $\mathrm{g}=20$, and propagation constant $k=-10$. (b) Profiles of the 1D flat-floor bubbles for different values of $\mathrm{g}$ at $x_{0}=5$ and $k=-8$. (c) Profiles of the same modes for different values of $k$ at $x_{0}=5$ and $\mathrm{g}=20$. (d) The power defect of the 1D flat-floor bubble, defined as Eq. (\protect\ref{DeltaP1D}), versus $k$, as obtained from the numerical solution (the black solid line), and as predicted by TFA, according to Eq. (\protect\ref{TF1D}) (the red dashed line), for the same $x_{0}=5$ and $\mathrm{g}=20$ as in (c). In Eq. (\protect\ref{gprofile}), $\mathrm{g}_{0}\equiv 1$ is fixed throughout this paper. The numerical scheme eith stepsizes $\Delta x=\Delta r=0.1$ and $\Delta z=0.002$ is used throughout this work, see details in the main text.}
\label{fig1}
\end{figure*}

The objective of this work is to extend the potential use of the spatially modulated self-repulsive nonlinearity. There are physically relevant species of nonlinear modes that have not been studied, in this context, in previous works. One of them represents \textquotedblleft bubbles", i.e., modes with a suppressed density in the central region, which, unlike spatially odd dark solitons, are spatially even states that do not cross zero. The bubbles were originally introduced in usual models with the uniform nonlinearity \cite{Barash}. Collisions between bubbles \cite{BUB1} and a possibility of collapse dynamics in them \cite{BUB2} were studied too. Models with spatially inhomogeneous self-defocusing, which is strong near the center and
weak at the periphery, are very natural settings for the study of bubbles.

Another important class of the modes is one with flat-top shapes. They have been studied, theoretically and experimentally, in many settings with competing nonlinearities, such as cubic-quintic optical media \cite{Cid} and \textquotedblleft quantum droplets", predicted in BEC with the attractive cubic nonlinearity balanced by the effective quartic self-repulsion induced by quantum fluctuations around mean-field states \cite{Petrov1,Petrov2}. The flat-top shape is explained by the fact that the competition imposes an upper limit on the density of the wave field, hence increase of the total norm of the field leads to the expansion of the mode, keeping its amplitude constant. In particular, the upper limit imposed on the density makes the BEC in the \textquotedblleft quantum droplets" an incompressible liquid, which explains the name of these states. The droplets have been created in the experiment \cite{Leticia1,Inguscio,hetero}, and their theoretical investigation was extended for states with embedded vorticity \cite{swirling,Raymond}.

Unlike the flat-top-shaped states, concepts of nonlinear modes whose shape would demonstrate features such as a \textquotedblleft flat floor" or \textquotedblleft flat waist" were not elaborated yet, to the best of our knowledge. In this paper, we explore such possibilities and demonstrate that 1D and 2D flat-floor bubbles, 1D flat-waist dark solitons and 2D flat-waist
vortex modes can be readily supported by inhomogeneous self-defocusing cubic nonlinearity, without the necessity to use composite competing nonlinearities. In addition to their interest to fundamental studies, these previously unexplored states may find applications, such as all-optical beam guiding \cite{Denz,Bliokh} (using the inner \textquotedblleft hole" in dark solitons, bubbles, or vortices as conduits steering embedded signal beams) and optical tweezers. In these contexts, the size of the hole is a crucially important parameter, which can be adjusted by means of the waist shaping.

The rest of the paper is organized as follows. The 1D and 2D models are formulated in Section \ref{sec2}, which also includes some analytical results, such as ones based on the Thomas-Fermi approximation (TFA) and the limit case of the model with the delta-functional spatial modulation of the local nonlinearity strength. Numerical results for bubbles, dark solitons (in 1D), and vortices (in 2D) are reported in Section \ref{sec3}. It also includes a qualitative analytical explanation for stability and instability of the dark solitons. The paper is concluded by Section \ref{sec4}.

\section{The model and methods}

\label{sec2}

\subsection{The basic equations and solutions}

\subsubsection{The flat-floor profile}

The nonlinear Schr\"{o}dinger (NLS) equation, which describes the propagation of laser beams in self-defocusing nonlinear media, or effectively 2D matter waves in BECs (in the latter context, it is called the Gross-Pitaevskii (GP) equation \cite{Gross,Pit0,Pit}), is taken in the scaled form:
\begin{equation}
i\frac{\partial E}{\partial z}=-\frac{1}{2}\nabla ^{2}E+G(r)\left\vert E\right\vert ^{2}E.
\label{NLSE}
\end{equation}
Here $E$ and $z$ denote the field amplitude and propagation distance, respectively, $(x,y)$ is the set of the transverse coordinates, with $r\equiv \sqrt{x^{2}+y^{2}}$, and $\nabla ^{2}=\partial_{x}^{2}+\partial _{y}^{2}$ is the paraxial-diffraction operator, in terms of optics, or the kinetic-energy operator for matter waves. In the latter case, $E$ and $z$ are replaced by wave function $\psi $ and time $t$, respectively. In the 1D version of Eq. (\ref{NLSE}), with single coordinate $x$, $G(r)$ is replaced by $G(x)$.

\begin{figure}[tbp]
\begin{center}
\includegraphics[width=1\columnwidth]{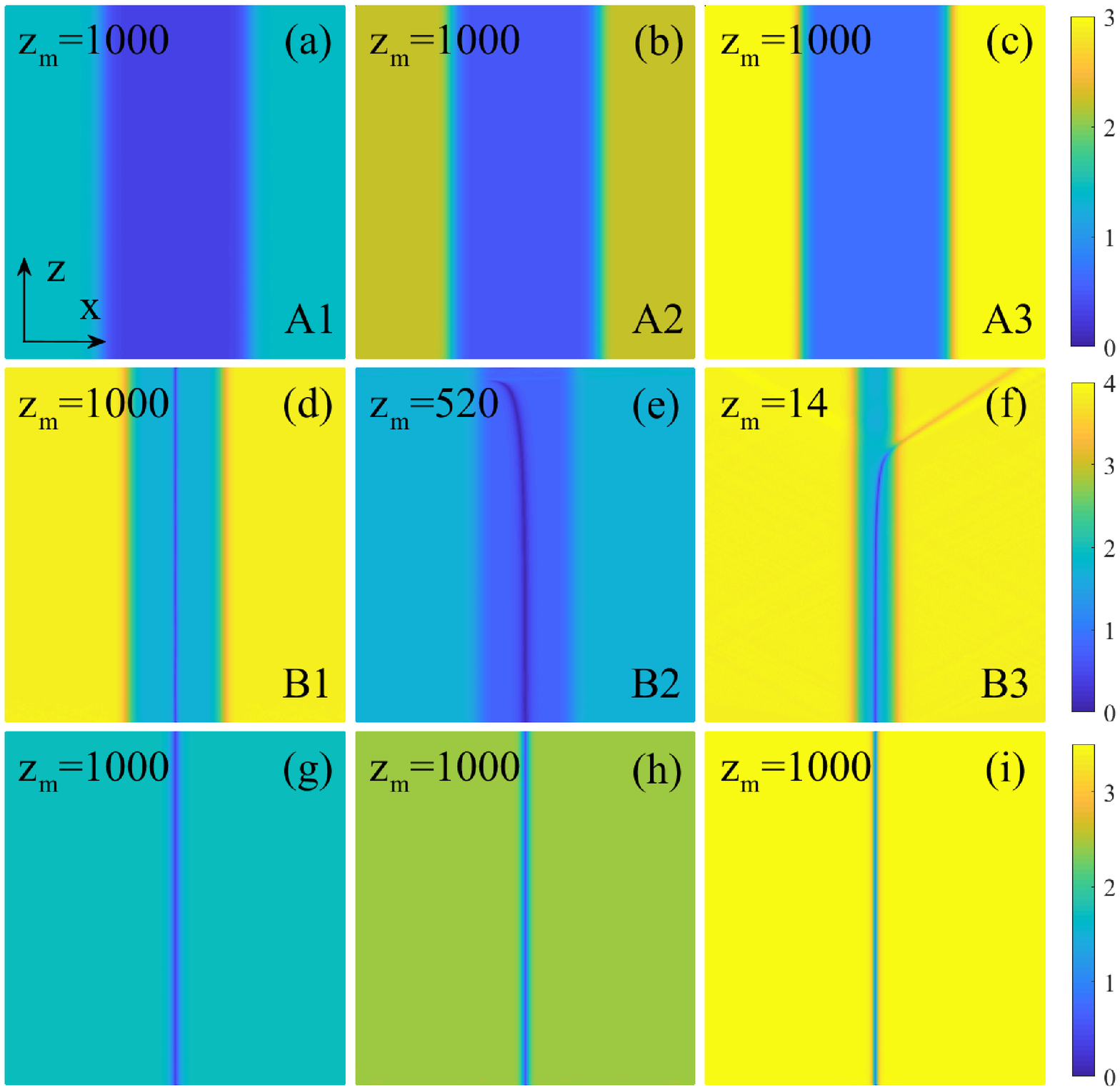}
\end{center}
\caption{Simulations of the stable propagation of perturbed 1D bubbles with different values of $k$ at $x_{0}=5$, $\mathrm{g}=20$: (a) $k=-2$; (b) $k=-5$; (c) $k=-9$. The propagation of perturbed 1D dark solitons at $\mathrm{g}=5$: (d) a stable dark soliton with $x_{0}=3$, $k=-15$; (e) an unstable one with $x_{0}=3$, $k=-3$; (f) an unstable dark soliton with $x_{0}=0.5$, $k=-15 $. Simulation of the stable propagations of perturbed 1D bubbles in the model based on Eq. (\protect\ref{delta}), with $\protect\gamma =225$ and the delta-function approximated by a narrow Gaussian, $\tilde{\protect\delta}(x)=\protect\sqrt{1000/\protect\pi }\mathrm{exp}(-x^{2}/0.001)$ (with $\protect\int_{-\infty }^{+\infty }\protect\delta(x)dx=1$, as it must be for
the delta-function): (g) $k=-3$; (h) $k=-6$; (i) $k=-12$. All panels display the domain with size $|x|\leq 15$, while $z_{\mathrm{m}}$ stands for the largest propagation distance in the simulations.}
\label{fig2}
\end{figure}

The NLS equation (\ref{NLSE}) conserves the total power (alias norm),
\begin{equation}
P=\iint dxdy\left\vert E\left( x,y\right) \right\vert ^{2},
\label{P}
\end{equation}
the angular momentum (in 2D, where it may be conveniently written in polar coordinates, $r$ and $\theta $),
\begin{equation}
M=i\int_{0}^{\infty }rdr\int_{0}^{2\pi }d\theta E\left( \partial E^{\ast}/\partial \theta \right).
\label{M}
\end{equation}
and the Hamiltonian,
\begin{equation}
H=\frac{1}{2}\iint dxdy\left[ \left\vert \nabla E\right\vert^{2}+G(r)|E\left\vert ^{4}\right\vert \right].
\label{H}
\end{equation}
In 1D, expressions (\ref{P}) and (\ref{H}) are replaced by their straightforward counterparts. Numerical simulations performed in this work maintain the conservation of $P$, $M$, and $H$.

\begin{figure*}[tbp]
\begin{center}
\includegraphics[width=1.9\columnwidth]{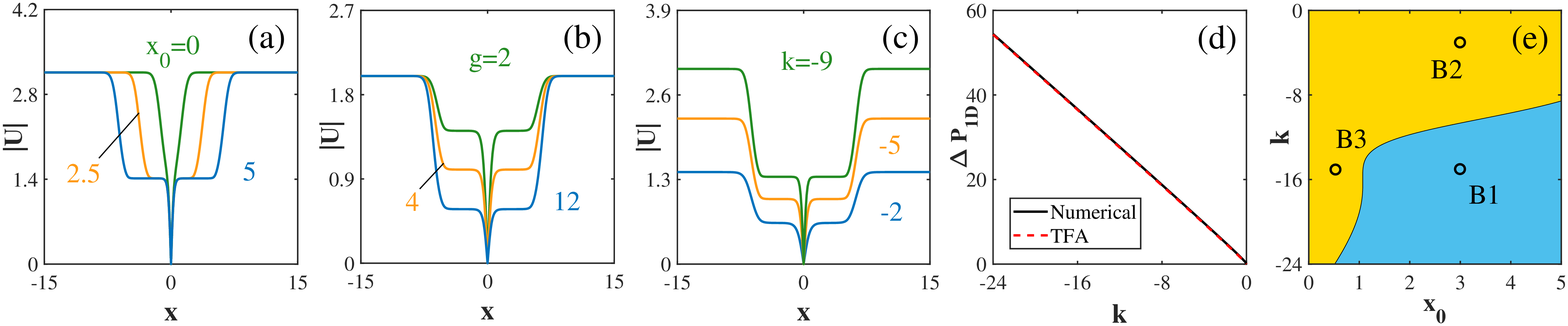}
\end{center}
\caption{(a) The transition of 1D ordinary dark solitons into flat-waist ones with the increase of the spatial-modulation width, $x_{0}=0,2.5,5$, at fixed $\mathrm{g}=5$ and $k=-10$. (b) Profiles of the 1D flat-waist dark solitons for different values of $\mathrm{g}$ at $x_{0}=5$ and $k=-4$. (c) Profiles of 1D flat-waist dark solitons with different values of $k$ at $%
x_{0}=5$ and $\mathrm{g}=5$. (d) The power defect of the 1D flat-waist dark solitons, given by Eq. (\protect\ref{DeltaP1D}), versus $k$, as obtained from the numerical solution (the black solid line), and as predicted by TFA, according to Eqs. (\protect\ref{TFDS}) (the red dashed line), for the same values $x_{0}=5$ and $\mathrm{g}=5$ as in (c). (e) Stability (blue) and
instability (yellow) domains for dark solitons in the plane of ($x_{0}$,$k$) at $\mathrm{g}=5$. Direct simulations of the propagations of the dark solitons marked by B1--B3 are displayed in Figs. \protect\ref{fig2}(d--f), respectively.}
\label{fig3}
\end{figure*}

To construct flat-floor bubbles and flat-waist vortices and dark solitons, we define the profile of the self-defocusing nonlinearity with $G(r)>0$ as
\begin{equation}
G(r)=\left\{
\begin{array}{c}
\mathrm{g},r\leq r_{0}, \\
\mathrm{g}_{0}+(\mathrm{g}-\mathrm{g}_{0})\mathrm{exp}[-(r-r_{0})^{2}],r>r_{0},
\end{array}
\right.  \label{gprofile}
\end{equation}
where $\mathrm{g}_{0}$, $r_{0}$, and $\mathrm{g}>\mathrm{g}_{0}$ are positive constants. In 1D, $r$ and $r_{0}$ should be replaced by $|x|$ and $x_{0}>0$. By means of rescaling, we set $\mathrm{g}_{0}\equiv 1$ in what follows below, while $\mathrm{g}>1$ represents the largest local strength of the self-defocusing nonlinearity, at $r=0$, and $r_{0}$ is the width of the
flat-floor section of the nonlinearity profile.

Stationary solutions with real propagation constant $k<0$ and embedded integer vorticity, $m=0,1,2,...$ , are sought for, in polar coordinates $\left( r,\theta \right) $, as
\begin{equation}
E=U(r)\mathrm{exp}(ikz+im\theta ),
\label{m}
\end{equation}
where the real stationary field amplitude $U$ satisfies the equation
\begin{equation}
kU-\frac{1}{2}\left( \frac{d^{2}}{dr^{2}}+\frac{1}{r}\frac{d}{dr}-\frac{m^{2}}{r^{2}}\right) U+G(r)U^{3}=0.
\label{NLSES}
\end{equation}
The mode with $m\geq 1$ may be considered as a vortex with winding number $m$ embedded in the fundamental state (the 2D \textquotedblleft bubble"), which corresponds to $m=0$.

The 1D stationary states are looked for as
\begin{equation}
E=U(x)\mathrm{exp}(ikz),
\label{1D}
\end{equation}
with real function $U$ satisfying the equation
\begin{equation}
kU-\frac{1}{2}\frac{d^{2}U}{dx^{2}}+G(x)U^{3}=0.
\label{U1D}
\end{equation}
The 1D counterpart of vortices is the dark soliton, which can be constructed by embedding the usual dark-soliton solution of the NLS equation with $G=\mathrm{const}$ \cite{Kivshar,Dimitri},
\begin{equation}
U_{\mathrm{dark}}(x)=\sqrt{-\frac{k}{G}}\tanh \left( \sqrt{-k}x\right) ,
\label{DS}
\end{equation}
in the flat-top segment of the 1D fundamental state. In the uniform space, the dark soliton is characterized by its \textit{power defect}, i.e., the difference of the integral power and its background value, corresponding to the constant asymptotic density, $U_{\mathrm{backgr}}^{2}=-k/G$:
\begin{equation}
\Delta P\equiv \int_{-\infty }^{+\infty }dx\left[ U_{\mathrm{backgr}}^{2}-U_{\mathrm{dark}}^{2}(x)\right] =\frac{2}{G}\sqrt{-k}.
\label{DeltaP}
\end{equation}
Another characteristic of the dark soliton is the difference of its Hamiltonian and the background value,
\begin{equation}
\begin{split}
\Delta H&\equiv \frac{1}{2}\int_{-\infty }^{+\infty }dx\left\{ G\left[ U_{\mathrm{backgr}}^{4}-U_{\mathrm{dark}}^{4}(x)\right]
-\left\vert \frac{\partial U}{\partial x}\right\vert ^{2}\right\} \\
&=\frac{2}{3G}\left( -k\right) ^{3/2}.
\label{DeltaH}
\end{split}
\end{equation}
For the dark soliton in the uniform space, it is relevant to express $\Delta H$ in terms of $\Delta P$, using Eqs. (\ref{DeltaP}) and (\ref{DeltaH}):
\begin{equation}
\Delta H=\frac{G^{2}}{12}\left( \Delta P\right) ^{3}.
\label{HP}
\end{equation}

Similarly to Eq. (\ref{DeltaP}), one can define the power defect for the bubble, vortex, and dark soliton states in the 2D and 1D versions of the present model, with the nonlinear-modulation profile defined as per Eq. (\ref{gprofile}). The defect is a sum of two terms, which represent the difference between the actual solution and background values in the regions of $r<r_{0}$ (or $|x|<x_{0}$, in 1D) and $r>r_{0}$ (or $|x|>x_{0}$):
\begin{equation}
\begin{split}
\Delta P_{\mathrm{2D}} &=2\pi \int_{0}^{r_{0}}rdr\left[ -\frac{k}{\mathrm{g}}-U^{2}(r)\right] \\
&+2\pi \int_{r_{0}}^{\infty }rdr\left[ -\frac{k}{\mathrm{g}_{0}}-U^{2}(r)\right] ,
\label{DeltaP2D}
\end{split}
\end{equation}
\begin{equation}
\begin{split}
\Delta P_{\mathrm{1D}} &=2\int_{0}^{x_{0}}dx\left[ -\frac{k}{\mathrm{g}}-U^{2}(x)\right] \\
&+2\int_{x_{0}}^{\infty }dx\left[ -\frac{k}{\mathrm{g}_{0}}-U^{2}(x)\right].
\label{DeltaP1D}
\end{split}
\end{equation}

\begin{figure*}[tbp]
\begin{center}
\includegraphics[width=1.6\columnwidth]{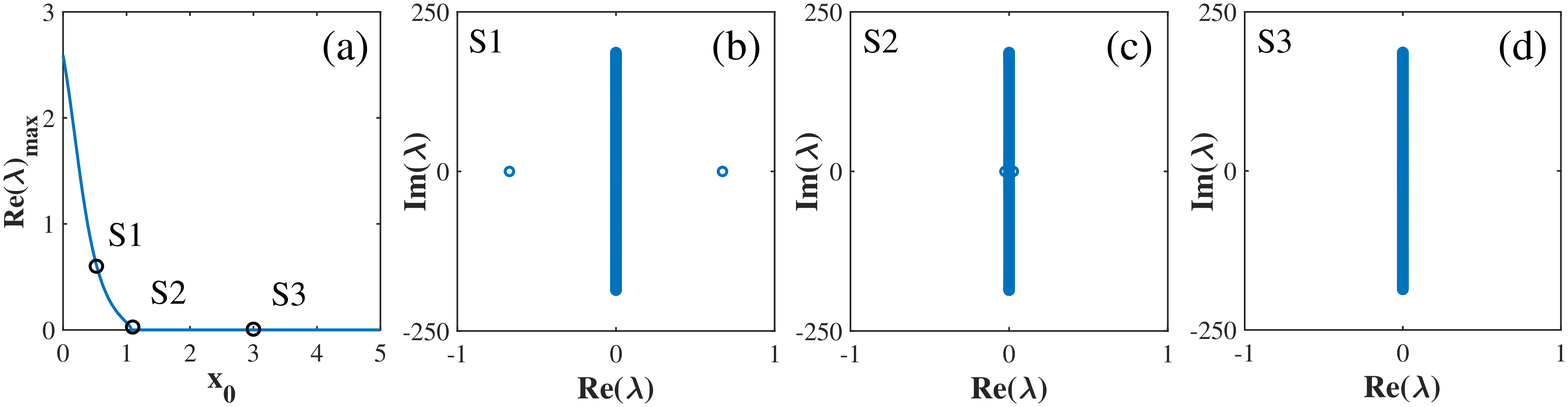}
\end{center}
\caption{(a) The largest real part of the eigenvalues, $\mathrm{Re(\protect\lambda )_{max}}$, defined as per Eq. (\protect\ref{pert}), versus $x_{0}$ for dark solitons at fixed $k=-15$, $\mathrm{g}=5$. Examples of the eigenvalue spectra at $k=-15$, $\mathrm{g}=5$, and values of $x_{0}$ marked by symbols S1--S3 in panel (a): $x_{0}=0.5$ in (b); $x_{0}=1.06$ in (c); $x_{0}=3$ in (d).}
\label{fig4}
\end{figure*}

\subsubsection{The delta-functional profile in 1D}

The limit case of the 1D model, opposite to the flat-floor configuration, is one with the central region that is very narrow in comparison with the dark-soliton's width. In this limit, Eq. (\ref{U1D}) is replaced by
\begin{equation}
kU-\frac{1}{2}\frac{d^{2}U}{dx^{2}}+\left[ 1+\gamma \delta (x)\right]U^{3}=0,
\label{delta}
\end{equation}
where $\delta (x)$ is the Dirac's delta-function, and the coefficient in front of it is $\gamma =\int_{-x_{0}}^{+x_{0}}G(x)dx\equiv 2\mathrm{g}x_{0}$, with $G(x)$ taken from the 1D version of Eq. (\ref{gprofile}). The interaction of field $U$ with the delta-functional defect is accounted for by the following term in the Hamiltonian (the interaction potential), cf. Eq. (\ref{DeltaH}):
\begin{equation}
W_{\mathrm{int}}=\left( \gamma /2\right) \left\vert U(x=0)\right\vert ^{4}.
\label{int}
\end{equation}

Note that Eq. (\ref{delta}) gives rise to an exact bubble solution,
\begin{equation}
U_{\mathrm{delta}}(x)=\sqrt{-k}\tanh \left( \sqrt{-k}\left( \left\vert x\right\vert +\xi \right) \right) ,
\label{Udelta}
\end{equation}
with offset $\xi $ determined by the cubic equation for $\tanh \left( \sqrt{-k}\xi \right) $:
\begin{equation}
\gamma \sqrt{-k}\tanh ^{3}\left( \sqrt{-k}\xi \right) +\tanh ^{2}\left(\sqrt{-k}\xi \right) -1=0.
\label{xi}
\end{equation}
It is easy to see that Eq. (\ref{xi}) has a single solution for all positive values of $\gamma $. The power defect of solution (\ref{Udelta}), defined as in Eq. (\ref{DeltaP}), is
\begin{equation}
\Delta P_{\mathrm{delta}}=2\sqrt{-k}\left[ 1-\tanh \left( \sqrt{-k}\xi \right) \right] .
\label{DeltaPdelta}
\end{equation}
Note that, as it follows from Eqs. (\ref{xi}) and (\ref{DeltaPdelta}), the power defect satisfies condition $d\Delta P/dk>0$, which is tantamount to the \textit{anti-Vakhitov-Kolokolov} criterion. It is a necessary condition for stability of states supported by self-repulsive nonlinearities \cite{anti}. This fact suggests that the exact bubbles supported by the
delta-functional nonlinearity-modulation profile may be stable (the Vakhitov-Kolokolov criterion proper, with the opposite sign, is a well-known necessary condition for the stability of localized states supported by self-attractive nonlinearities \cite{VK,Berge,Fibich}). Indeed, direct simulations of perturbed evolution of these bubbles, performed in the framework of Eq. (\ref{NLSE}) with $G(x)=\gamma \tilde{\delta}(x)$ (see Eq. (\ref{delta})), where $\tilde{\delta}(x)$ is a narrow Gaussian approximating the ideal delta-function, corroborate the stability, as shown below in Figs. \ref{fig2}(g-i).

\subsection{Linearized equations for small perturbations}

To explore the stability of the stationary solutions against small perturbations, we introduce the perturbed 1D solution as
\begin{equation}
E=[U(x)+p(x)\mathrm{exp}(\lambda z)+q^{\ast }(x)\mathrm{exp}(\lambda ^{\ast}z)]\mathrm{exp}(ikz),
\label{pert}
\end{equation}
where $p(x)$ and $q^{\ast }(x)$ are small perturbations associated with an instability growth rate, $\lambda $. Substituting this ansatz in Eq. (\ref{NLSE}) leads to the eigenvalue problem for $\lambda $,
\begin{eqnarray}
i\lambda p &&=-\frac{1}{2}\frac{d^{2}p}{dx^{2}}+kp+\mathrm{g}U^{2}(2p+q),
\notag \\
i\lambda q &&=+\frac{1}{2}\frac{d^{2}q}{dx^{2}}-kq-\mathrm{g}U^{2}(2q+p).
\label{LAS1D}
\end{eqnarray}

In the 2D case, the perturbed wave function with embedded integer vorticity, $m$, and an integer azimuthal index, $n$, of the perturbation is taken as
\begin{equation}
E=[U(r)+p(r)e^{in\theta +\lambda z}+q^{\ast }(r)e^{-in\theta +\lambda ^{\ast}z}]e^{im\theta +ikz},
\end{equation}
which leads to the eigenvalue problem in the following form:
\begin{eqnarray}
i\lambda p =&&-\frac{1}{2}\left[ \frac{d^{2}}{dr^{2}}+\frac{1}{r}\frac{d}{dr}-\frac{(m+n)^{2}}{r^{2}}\right] p+kp  \notag \\
&&+\mathrm{g}U^{2}(2p+q),  \notag \\
i\lambda q =&&+\frac{1}{2}\left[ \frac{d^{2}}{dr^{2}}+\frac{1}{r}\frac{d}{dr}-\frac{(m-n)^{2}}{r^{2}}\right] q-kq  \notag \\
&&-\mathrm{g}U^{2}(2q+p).
\label{LAS2D}
\end{eqnarray}
Stationary perturbed solutions are stable if real parts of all the corresponding eigenvalues are zero, $\mathrm{Re}(\lambda )=0$.

\subsection{The Thomas-Fermi approximation (TFA)}

\begin{figure*}[tbp]
\begin{center}
\includegraphics[width=1.6\columnwidth]{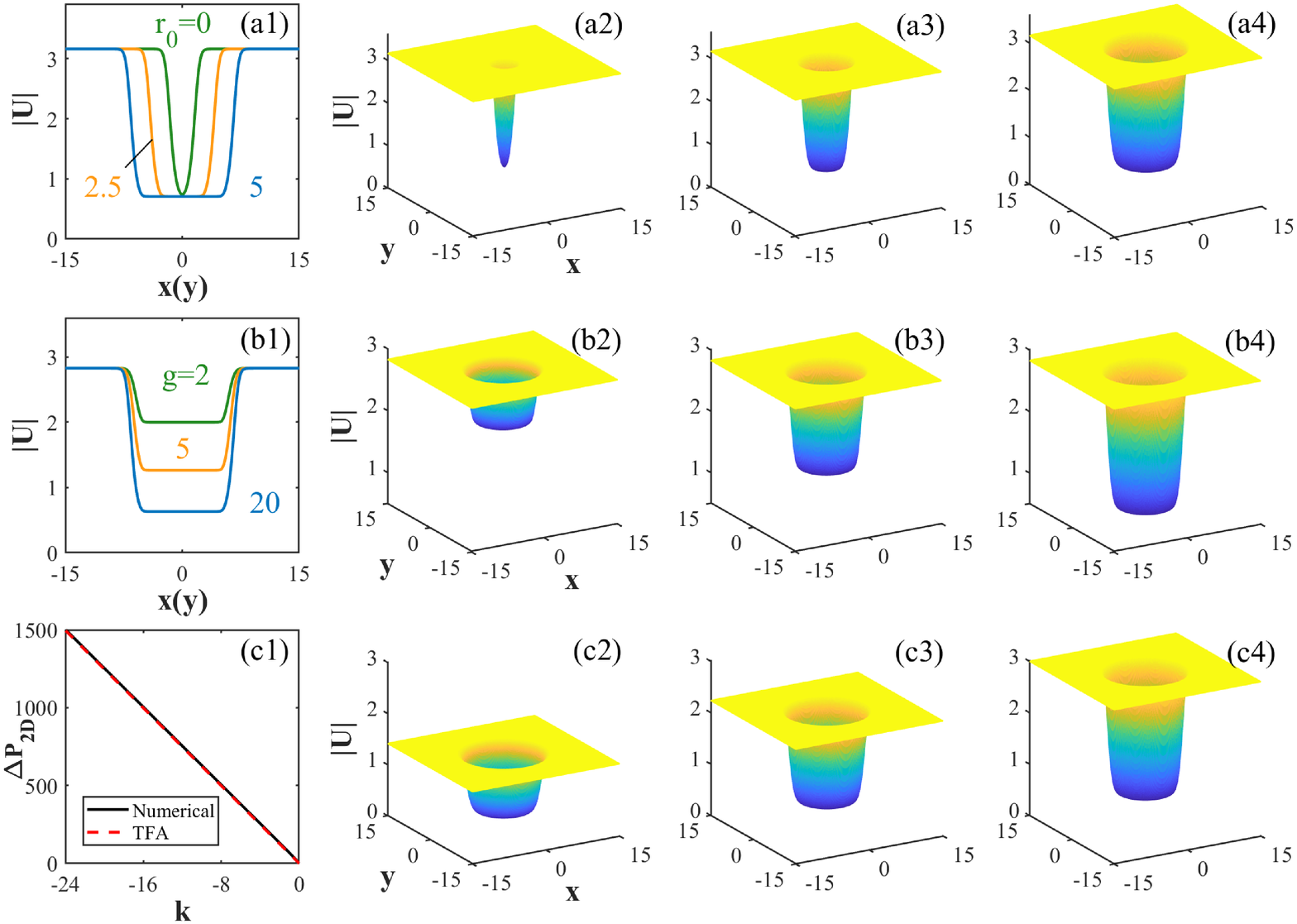}
\end{center}
\caption{Cross-section profiles and full intensity distributions for 2D bubbles at $\mathrm{g}=20$, $k=-10$: (a1) profiles obtained for different values of $r_{0}$ in Eq. (\protect\ref{gprofile}); (a2) the intensity distribution for $r_{0}=0$; (a3) the distribution for $r_{0}=2.5$; (a4): the same for $r_{0}=5$. The cross-section profiles and full distributions of the intensity for 2D flat-floor bubbles at $r_{0}=5$, $k=-8$: (b1) the profiles for different values of $\mathrm{g}$; (b2) the intensity profile for $\mathrm{g}=2$; (b3) the profile for $\mathrm{g}=5$; (b4): the same for $\mathrm{g}=20$. (c1) The power defect of the 2D flat-floor bubble, defined as per Eq. (\protect\ref{DeltaP}), versus $k$, as obtained from the numerical results (the black solid line) and TFA (the red dashed line). The full intensity distributions for the 2D flat-floor bubbles, obtained for $r_{0}=5$, $\mathrm{g}=20$: (c2) for $k=-2$; (c3) for $k=-5$; (c4) for $k=-9$. All bubbles are dislayed as real profiles, as their phases are constant.}
\label{fig5}
\end{figure*}

TFA, which neglects the derivatives in Eq. (\ref{NLSES}), is a commonly known method for constructing stationary solutions \cite{Pit}. In the framework of TFA, the approximate solution of Eq. (\ref{NLSES}) is
\begin{equation}
\left[ U_{\mathrm{TFA}}^{2}(x)\right] _{\mathrm{1D}}=-\frac{k}{G(x)},
\label{TF1D}
\end{equation}
\begin{equation}
\left[ U_{\mathrm{TFA}}^{2}(r)\right] _{\mathrm{2D}}=\left\{
\begin{array}{c}
-\left( k+m^{2}r^{-2}\right) /G(r),~\mathrm{at}~r^{2}>-m^{2}/k, \\
0,~\mathrm{at}~r^{2}<-m^{2}/k~.
\end{array}
\right.
\label{TF2D}
\end{equation}
Obviously, Eq. (\ref{TF1D}) pertains to the bubble, while the 2D expression (\ref{TF2D}) predicts a 2D bubble for both $m=0$ and vortices with $m\geq 1$ (in the former case, the \textquotedblleft hole" at $r^{2}<-m^{2}/k$ is absent in the TFA solution). Note also that, for the radial-modulation profile (\ref{gprofile}), the inner flat area, $r<r_{0}$, is completely covered by the zero part of TFA, see the bottom row in Eq. (\ref{TF2D}), for large values of the vorticity,
\begin{equation}
m^{2}>-kr_{0}^{2}.
\label{m^2}
\end{equation}
The predictions produced by TFA for the 1D and 2D bubbles and vortices are compared below with numerical solutions.

In 1D, the TFA does not directly produce dark solitons, but they can be introduced by means of a straightforward approximation combining Eqs. (\ref{DS}) and (\ref{TF1D}),
\begin{equation}
\left[ U_{\mathrm{TFA}}(x)\right] _{\mathrm{dark}}=\sqrt{-\frac{k}{G(x)}}\tanh \left( \sqrt{-k}x\right),
\label{TFDS}
\end{equation}
which is relevant under condition $\sqrt{-k}x_{0}\gg 1$.

The accuracy of TFA can be naturally characterized by calculating the power defects, defined by Eqs. (\ref{DeltaP2D}) and (\ref{DeltaP1D}), and comparing them to counterparts produced by the numerical solution of Eqs. (\ref{NLSES}) and (\ref{U1D}). In particular, it is obvious from Eqs. (\ref{TF1D}) and (\ref{TF2D}) (with $m=0$) that TFA predicts the power defect of
the 1D and 2D bubbles to be proportional to $|k|$. For the 2D state with $m=0 $, it can be calculated analytically in the limit of $r_{0}=0$:
\begin{equation}
\Delta P_{\mathrm{2D}}\left( m=0,r_{0}=0\right) =-\frac{\pi k}{\mathrm{g}_{0}}\ln \left( \frac{\mathrm{g}}{\mathrm{g}_{0}}\right).
\label{r_0=0}
\end{equation}
For the 1D bubbles, TFA produces a simple result in the limit of $\mathrm{g}\gg 1$,
\begin{equation}
\Delta P_{\mathrm{1D}}\approx -\frac{2k}{\mathrm{g}_{0}}\sqrt{\ln \left(\frac{\mathrm{g}}{\mathrm{g}_{0}}\right) }.
\label{g>>1}
\end{equation}

The comparison of the TFA-predicted and numerically found $\Delta P_{\mathrm{1D,2D}}$ is displayed below in Figs. \ref{fig1}(d) and \ref{fig5}(c1), respectively.

\section{Results and discussion}

\label{sec3}
The numerical solution of Eqs. (\ref{NLSES}) and (\ref{U1D}) was produced by means of the Newton's method. It is an example of root-finding algorithms for $f(x)=0$, solved by means of the iteration process, $x_{n+1}=x_{n}-f(x_{n})/f^{\prime }(x_{n})$ where $n$ denotes the iteration's number and $f^{\prime }$ stands for the first derivative \cite{Newton}. In the case of the 2D equation (\ref{NLSES}), the method was applied in the polar coordinates, removing the artificial singularity at $r\rightarrow 0$ by means of the appropriate boundary conditions, \textit{viz}., $dU/dr|_{r=0}=0$ for zero-vorticity modes, or $U(r)\sim r^{m}$ ones with vorticity $m\geq 1$. The input for these modes was taken as $U_{\mathrm{input}}=\sqrt{-k/G}\left( \mathrm{tanh}(\sqrt{-k}r)\right) ^{m}$, replacing $r$ by $x$ in 1D, and setting $m=1$ for the 1D dark solitons.

The stability of the stationary solutions was subsequently explored by solving eigenvalue equations (\ref{LAS1D}) and (\ref{LAS2D}), and the results were verified by direct simulations of Eq. (\ref{NLSE}) for the perturbed propagation, employing the finite-difference method for marching in time \cite{FDTD2}. It was checked that the numerical mesh with $\Delta
x=\Delta r=0.1$ and $\Delta z=0.002$ was sufficient for producing fully reliable results (i.e., rerunning the calculations with essentially smaller $\Delta x,\Delta r$ and $\Delta z$ does not change the findings). It is relevant to mention that numerical methods for producing stationary and dynamical solutions to NLS/GP equations with a spatially varying nonlinearity coefficients were developed and used in many previous works \cite{review,GAP13,anti,NL2,CQNL,NL3}, \cite{SDN1}-\cite{SDN13}.

\subsection{One-dimensional (1D) bubbles}

\begin{figure*}[tbp]
\begin{center}
\includegraphics[width=1.7\columnwidth]{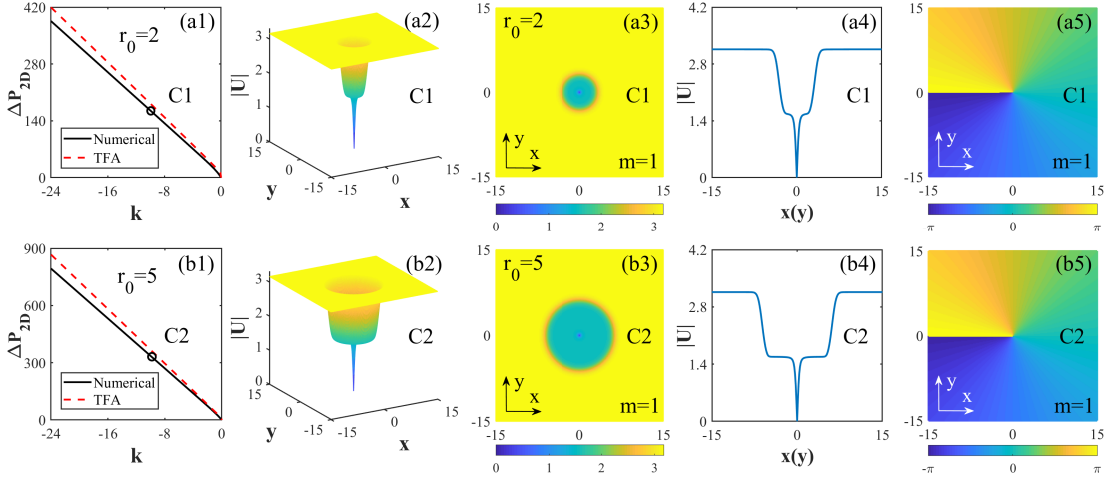}
\end{center}
\caption{Characteristics of flat-waist vortices with $m=1$, $\mathrm{g}=4$, and different values of $r_{0}$. (a1,b1): The power defect, defined as per Eq. (\protect\ref{DeltaP}), versus $k$. Other panels represent typical examples of the vortices with $k=-10$, by showing their density distributions in (a2,b2), contour plots (a3,b3), cross-section profiles (a4,b4), and phase patterns (a5,b5). The values of $r_{0}$ are: $2$ in (a1)-(a5), and $5$ in (b1)-(b5).}
\label{fig6}
\end{figure*}

Typical profiles of the flat-floor bubbles in 1D are displayed in Fig. \ref{fig1}. The transition of the ordinary bubbles into ones featuring the flat-floor profile, following the increase of width $x_{0}$ of the flat section in the nonlinearity profile, is displayed in Fig. \ref{fig1}(a). Naturally, the width of the flat segment in the bubble is determined by $x_{0}$. The width is not sensitive to the strength of the nonlinearity in the flat section, $\mathrm{g}$, while the minimum value of the field in the bubble mode depends on $\mathrm{g}$, as seen in Fig. \ref{fig1}(b). Profiles of the flat-floor bubbles with different values of the propagation constant, $k$, are compared in Fig. \ref{fig1}(c).

The power defect versus propagation constant $k$ of 1D flat-floor bubbles, as obtained from the numerical results and as predicted by TFA, are shown in Fig. \ref{fig1}(d). Note that the TFA and numerical findings are virtually identical, with $\Delta P_{\mathrm{1D}}$ growing linearly with $|k|$, in exact agreement with TFA. Furthermore, for $\mathrm{g}=20$ and $\mathrm{g}_{0}=1$, which corresponds to Fig. \ref{fig1}(d), Eq. (\ref{g>>1}) predicts the slope $d\left( \Delta P_{\mathrm{1D}}\right) /dk=$ \ $2\sqrt{\ln 20}\approx \allowbreak 3.46$, which is very close to the value $\allowbreak 3.38 $ produced by Fig. \ref{fig1}(d).

The calculation of stability eigenvalues for the flat-floor bubbles has demonstrated that they are completely stable (at least, up to $k=-24$). This conclusion is corroborated by direct simulations of the modes marked by A1--A3 in Fig. \ref{fig1}(d), as shown in Figs. \ref{fig2}(a-c), respectively.

\begin{figure*}[tbp]
\begin{center}
\includegraphics[width=2\columnwidth]{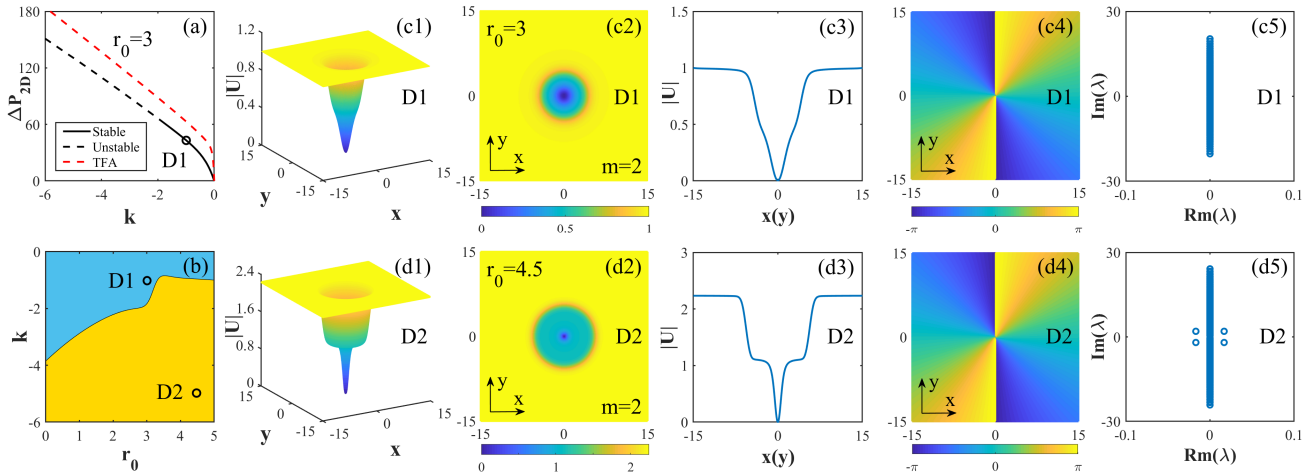}
\end{center}
\caption{Characteristics of double vortices, with $m=2$ and $\mathrm{g}=4$. (a) The power defect versus $k$ for a family of vortices with $r_{0}=3$. Both the numerical results and TFA prediction, calculated as per Eq. (\protect\ref{TF2D}), are displayed by the continuous black and dashed red lines, respectively. (b) Stability (blue) and instability (yellow) domains
for the double vortices in the plane of $(r_{0},k)$. Perturbed evolution of the vortices marked by D1 and D2 is displayed in Figs. \protect\ref{fig8}(c) and (d), respectively. Panels (c1)-(c5) display an example of a stable double vortex from the family, for $k=-1$. (c1): The intensity distribution; (c2): the contour plot; (c3): the cross-section profile; (c4): the phase pattern; (c5): the spectrum of stability eigenvalues. Panels (d1)-(d5) display the same as, respectively, (c1)-(c5), but for an unstable double vortex with $r_{0}=4.5$ and $k=-5$. Note that, as seen in panel (d5), the splitting instability of the double vortices (in the case when they are subject to the instability) is accounted for by the \textit{quartet} of
critical eigenvalues of perturbation modes, in the form given by Eq. (\protect\ref{quartet}).}
\label{fig7}
\end{figure*}

\subsection{Dark solitons}

Depicted in Fig. \ref{fig3} are profiles and the stability area of 1D dark solitons, constructed with the help of the input given by Eq. (\ref{TFDS}). Profiles of the solitons with different values of width $x_{0}$ of the flat section in the nonlinearity profile are shown in Fig. \ref{fig3}(a), where one can see that the ordinary dark solitons transform into flat-waist ones as $x_{0}$ increases. We present the profiles of the 1D dark solitons with different values of nonlinearity strength $\mathrm{g}$ in Fig. \ref{fig3}(b), in which the amplitude of the flat-waist segment decreases with the increase of $\mathrm{g}$. Then, the profiles of dark solitons with different propagation constants $k$ are shown in Fig. \ref{fig3}(c). Note that the variation of the dark-soliton profiles following the change of $x_{0}$, $\mathrm{g}$, and $k$, observed in Fig. \ref{fig3}, are quite similar to those for the 1D bubbles shown in Fig. \ref{fig1}.

Properties of the dark-soliton families are summarized in Fig. \ref{fig3}(d) by means of lines showing the relation between the soliton's power defect and propagation constant. It is seen that the TFA, based on Eqs. (\ref{TFDS}), produces the $\Delta P(k)$ dependence for the dark solitons which is virtually identical to its numerically computed counterpart.

An essential result is reported in Fig. \ref{fig3}(e): the stability domain for the 1D dark solitons in the ($x_{0}$,$k$) plane, produced by the solution of the eigenvalue problem based on Eq. (\ref{LAS1D}). It is seen that the decrease of the width of the flat-floor area, $x_{0}$, or the decrease of $|k|$ lead to destabilization of the dark soliton. In either case, it tends to lose its stability when it becomes relatively wide, in comparison with width $2x_{0}$ of the central region. Further, Fig. \ref{fig4} demonstrates that the destabilization, which occurs with the decrease of $x_{0}$ at a fixed value of $k_{0}$, is accounted for by a simple bifurcation of the center-saddle type \cite{Champneys}, with two imaginary eigenvalues colliding at the critical point and carrying over into a pair of real ones with opposite signs.

The stability of sufficiently narrow dark solitons can be understood following the principle that the stable configuration is one minimizing the system's Hamiltonian. To implement this approach, it is necessary to compare values of $\Delta H$, given by Eq. (\ref{DeltaH}), for the dark soliton placed in the central region, with $G=\mathrm{g}$, and in the peripheral
area ($\left\vert x\right\vert >x_{0}$), with $G=1$, taking into regard the constraint that this should be done for equal values of the total power. Indeed, the values of the power defect, given by Eq. (\ref{DeltaP}), are different for the same $k$ but different local values of $G(x)$, the defect being larger for smaller $G=1$ than for $G=\mathrm{g}$:
\begin{equation}
\delta P\equiv \Delta P\left( G=1\right) -\Delta P\left( G=\mathrm{g}\right)=2\sqrt{-k}\left( 1-\frac{1}{\mathrm{g}}\right).
\label{deltaP}
\end{equation}
To maintain the balance of the total power, one may assume that the imbalance, $\delta P$, is distributed in a very broad flat background, with large spatial width $L$, so that the constant background amplitude, $U_{\mathrm{backr}}=\sqrt{-k}$, for $G=1$ (see Eq. (\ref{TF1D})), is replaced by $U_{\mathrm{backr}}=\sqrt{-k}+\delta U$, with a small change,
\begin{equation}
\delta U\approx \delta P/\left( 2\sqrt{-k}L\right),
\label{deltaU}
\end{equation}
which absorbs the imbalance of the total power. This shift, in turn, gives rise to a small change in the quartic term of Hamiltonian (\ref{H}) in the background area:
\begin{equation}
\delta H_{\mathrm{backr}}\approx 2GU_{\mathrm{backr}}^{3}\delta U\cdot L\approx (-k)\delta P,
\label{deltaH}
\end{equation}
where expression (\ref{deltaU}) is substituted for $\delta U$. Finally, the total difference of values of the Hamiltonian for the dark soliton with the center placed at $x=0$ and at $\left\vert x\right\vert >x_{0}$ is
\begin{equation}
\begin{split}
\delta H_{\mathrm{total}}&=\delta H_{\mathrm{backr}}+\Delta H\left( G=\mathrm{g}\right) -\Delta H\left( G=1\right) \\
&\equiv \frac{4}{3}\left( -k\right) ^{3/2}\left( 1-\frac{1}{\mathrm{g}}\right).
\label{deltaHtotal}
\end{split}
\end{equation}

The positiveness of $\delta H_{\mathrm{total}}$ in Eq. (\ref{deltaHtotal}) suggests that the configuration with a relatively narrow dark soliton, placed in the broad top-floor region, with a larger value of $G=\mathrm{g}$, is stable, in comparison with its counterpart in which the dark soliton is placed in the asymptotic area, where the local nonlinearity coefficient, $%
G=1 $, is smaller. Note a crucially important contribution of term (\ref{deltaH}), which represents the change of the Hamiltonian of the indefinitely broad background due to the small change of its amplitude: without this term, $\delta H_{\mathrm{total}}$ would have a wrong sign.

On the other hand, for smaller $|k|$ or smaller values of $x_{0}$, when the dark soliton does not remain very narrow in comparison with size $x_{0}$ of the flat-floor region, Fig. \ref{fig3}(e) demonstrates that a relatively wide dark soliton, with the center placed at $x=0$, is unstable. This fact may be qualitatively explained, considering the limit case of the very narrow modulation profile approximated by Eq. (\ref{delta}). Indeed, treating the dark soliton as a quasi-particle in the framework of the perturbation theory \cite{pert}, the Hamiltonian perturbation term (\ref{int}) acts on the particle as an effective potential, obtained by the substitution of the unperturbed dark soliton (\ref{DS}), with its center
placed at $x=\xi $:
\begin{equation}
W_{\mathrm{int}}(\xi )=\left( \gamma /2\right) \tanh ^{4}\left( \sqrt{-k}\xi \right).
\label{W}
\end{equation}
By itself, this potential is attractive, with a minimum at $\xi =0.$ However, the effective dynamical mass of the dark soliton, treated as a particle, is \emph{negative} \cite{Kivshar,Dimitri,Busch}, therefore the action of potential (\ref{W}) makes the equilibrium position of the particle at $\xi =0$ unstable.

The predictions for the (in)stability of dark solitons based on the computation of eigenvalues are confirmed by direct simulations of their perturbed evolution. Typical examples, corresponding to points marked B1--B3 in Figs. \ref{fig3}(e), are displayed in Figs. \ref{fig2}(d--f), respectively. In particular, if the position of dark soliton placed at the center is unstable, in panels (e) and (f) of the Fig. \ref{fig2}, it is spontaneously expelled from the flat-floor area, as might be expected.

\begin{figure}[tbp]
\begin{center}
\includegraphics[width=1\columnwidth]{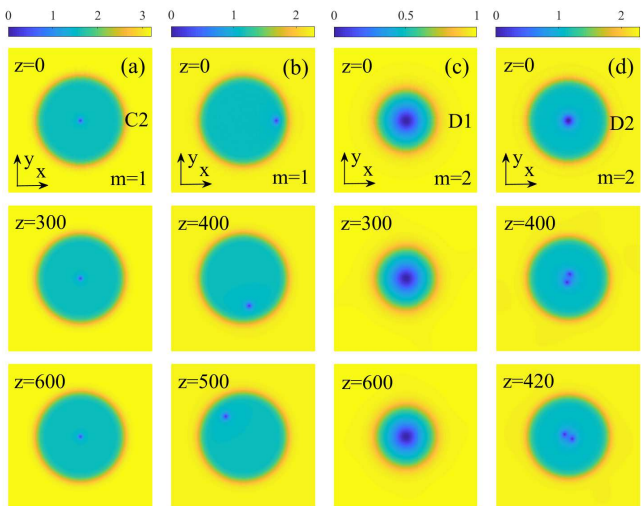}
\end{center}
\caption{The evolution of perturbed vortex states at $\mathrm{g}=4$: (a) the one with $m=1$, $r_{0}=5$, $k=-10$; (b) precession (circular motion) of the vortex with the pivot initially placed off the center, at $(x_{c},y_{c})=(4.7,0)$, with $m=1$, $r_{0}=5$, $k=-5$; (c) a stable double vortex with $m=2$, $r_{0}=3$, $k=-1$; (d) an unstable double vortex with $m=2 $, $r_{0}=4.5$, $k=-5$. All panels display the domain with size $|x|,|y|\leq 10$.}
\label{fig8}
\end{figure}

\subsection{Two-dimensional (2D) modes: bubbles and vortices}

We now turn to the consideration of 2D flat-floor bubbles (which represent the ground state) and flat-waist vortices. At first, we dwell on results for the bubbles. The profiles and intensity distributions for them are displayed in Fig. \ref{fig5}. In particular, Fig. \ref{fig5}(a1) depicts the profiles of 2D flat-floor bubbles for different values of $r_{0}$ in the
spatial-modulation pattern given by Eq. (\ref{gprofile}), whose intensity distributions are presented in Figs. \ref{fig5}(a2-a4). It is seen that the bubbles form the flat-floor shape as $r_{0}$ increases. Similarly, Fig. \ref{fig5}(b1) depicts the profiles of 2D flat-floor bubbles for different values of $\mathrm{g}$, whose 2D intensity distributions are shown in Figs. %
\ref{fig5}(b2-b4). In Figs. \ref{fig5}(b1--b4) one sees that the minimum value of the bubble's field naturally decreases with the increase of $\mathrm{g}$, in accordance with Eq. (\ref{TF2D}). Generally, the profiles of these 2D flat-floor bubbles are similar to their 1D counterparts, cf. Fig. \ref{fig1}. Further, according to the results of the linear stability
analysis and numerical simulations, the flat-floor bubbles are all stable (as one may expect for the ground states, which are actually represented by the bubbles), at least up to $|k|=24$. Dependencies of the power defect, $\Delta P_{\mathrm{2D}}$, on $k$ for the 2D flat-floor bubbles, as produced by the numerical results and TFA (see Eq. (\ref{TF2D})), are reported in
Fig. \ref{fig5}(c1), and the intensity distributions with different $k$ are depicted in Figs. \ref{fig5}(c2-c4). Note that the numerically found and TFA-predicted $\Delta P_{\mathrm{2D}}(k)$ dependencies are virtually identical, both being proportional to $|k|$, as said above (see, in particular, Eq. (\ref{r_0=0})).

\begin{figure*}[tbp]
\begin{center}
\includegraphics[width=2\columnwidth]{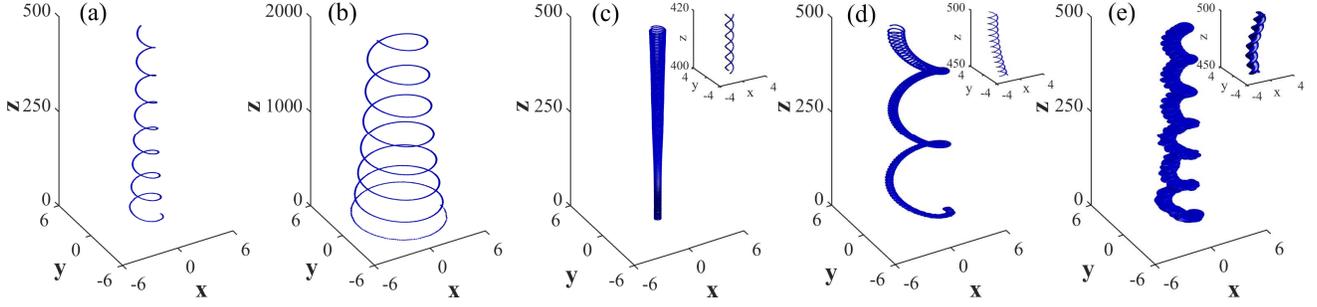}
\end{center}
\caption{Panels (a) and (b) display the precession (shown as the trajectory of motion of the vortex' pivot) of vortices with $m=1$, $\mathrm{g}=4$, $k=-5 $: (a) the vortex with the pivot initially placed at $(x_{c},y_{c})=(1.5,0)$, for $r_{0}=2$; (b) the vortex with $(x_{c},y_{c})=(4.7,0)$, for $r_{0}=5$. Panel (c) displays the trajectories of two unitary vortices, into which the unstable double one (with $m=2$) splits, for $\mathrm{g}=4$, $r_{0}=4.5$, $k=-5$. The precession of double vortices with $\mathrm{g}=4$: (d) an initially unstable one, with $r_{0}=5$, $k=-5$, $(x_{c},y_{c})=(3,0)$; (e) an initially stable double vortex, with $r_{0}=3$, $k=-2$, $(x_{c},y_{c})=(1.5,0)$. Insets in panels (c)-(e) show details of relatively short segments of the respective trajectories.}
\label{fig9}
\end{figure*}

Next we present the results for flat-waist vortices. Typical examples of such modes with winding number $m=1$ are displayed in Fig. \ref{fig6} for $r_{0}=2$ and $r_{0}=5$. The power defect versus $k$ for the 2D flat-waist vortices, as produced by the numerical solution and predicted by and TFA (see Eq. (\ref{TF2D}) for different values of $r_{0}$, is displayed in Figs. \ref{fig6}(a1,b1).\ It is seen that, unlike the 2D bubbles (cf. Fig. \ref{fig5}(c1)), there is a small discrepancy between the numerical findings and TFA prediction, and the power defect is not strictly proportional to $|k| $. Typical intensity distributions, contour plots, cross-section profiles, and phase patterns of the vortices are presented in Figs. \ref{fig6}(a2-a5) and \ref{fig6}(b2-b5). These modes naturally feature flat waists instead of the flat floor, as the vorticity stipulates vanishing of the amplitude at $r=0$. The width of the flat waist of the vortices increases with $r_{0}$, similar to the 2D bubble (ground-state) modes presented in Fig. \ref{fig5}. The phase patterns are usual, directly corresponding to $m=1$.

The results for double flat-waist vortices, with the winding number $m=2$, are depicted in Fig. \ref{fig7}. The dependence of the power defect on $k$, intensity distributions, contour plots, and cross-section profiles of such double vortices are similar to those with $m=1$, while the phase patterns in panels \ref{fig7}(c4) and (d4) obviously represent the double vorticity. Note a visible discrepancy between the numerical findings and TFA prediction for the power defect, as a function of $k$, in panel (a1), similar to what is shown for the 2D bubbles in Fig. \ref{fig6}(b1).

An essential issue is the stability of the vortices. According to our results, obtained from the linear-stability analysis and corroborated by direct simulations, the vortices with $m=1$ are completely stable, at least up to $k=-24$. On the other hand, the double vortices are stable only if $|k| $ is not too large. A stability chart for them in the plane of ($r_{0},k$) is shown in Fig. \ref{fig7}(b).

Simulations of the perturbed evolution of vortices from Figs. \ref{fig6} and \ref{fig7} are displayed in Fig. \ref{fig8}. First, panels (a) and (c) of this figure demonstrate the stable propagation of the unitary and double vortices labeled C2 in Fig. \ref{fig6} and D1 in Fig. \ref{fig7}, respectively.

Further, Fig. \ref{fig8}(b) displays \textit{precession} of the vortex whose pivot (center), with coordinates $(x_{c},y_{c})=\left( 4.7,0\right) $, is initially shifted off the origin. In Fig. \ref{fig8}(b), the pivot performs a circular motion in the periphery of the flat-waist area of the vortex. The trajectory of the motion is shown in Fig. \ref{fig9}(b). In addition, Fig. \ref{fig9}(a) shows the precession trajectory of the unitary vortex with the initial position $\left( x_{c},y_{c}\right) =\left( 1.5,0\right) $. In the latter case, the radius of the precession motion decreases in the course of the evolution. A general conclusion is that, if the original vortex is stable, its precessing motion is robust as well. These results resemble
experimentally observed precession of vortices in BEC \cite{precession}.

Unstable evolution of the vortex labeled D2 in Fig. \ref{fig7} is displayed in Figs. \ref{fig8}(d) and \ref{fig9}(c), where the double vortex (with $m=2$) splits into a persistently rotating pair of unitary ones, with the rotation period $z=7.39$. The character of the instability corresponds to the spectrum of eigenvalues for small perturbations around the double-vortex solution, which is displayed in Fig. \ref{fig7}(d5). It is seen, in this figure, that the instability is accounted for by a \textit{quartet} of eigenvalues with nonzero imaginary parts,
\begin{equation}
\lambda =\pm \text{Re}(\lambda )\pm i\text{Im}(\lambda )
\label{quartet}
\end{equation}
(with mutually independent pairs of $\pm $), on the contrary to the pair of purely real eigenvalues that represent the instability of the 1D dark solitons in Figs. \ref{fig4}(a,b). The imaginary part of the unstable eigenvalues, i.e., Im$(\lambda )$ in Eq. (\ref{quartet}), determines the angular velocity of the rotation of the emerging pair of split unitary vortices in Figs. \ref{fig8}(d) and \ref{fig9}(c). The same mechanism, dominated by the quartet of critical eigenvalues, accounts for the instability of the double vortices in the entire yellow (unstable) area in Fig. \ref{fig7}(b). Actually, the emergence of a quartet of complex eigenvalues is known as a generic type of the destabilizing bifurcation. In particular, it typically accounts for the onset of instability of binary vortices trapped in an external potential \cite{Skryabin} and discrete vortex solitons \cite{discr-vort}.

Actually, the splitting instability of double vortices (without setting the emerging pair in rotation) is a commonly known property of the 2D NLS equation with the spatially uniform cubic self-repulsion, see review \cite{Fetter} and original works \cite{Neu,Pu,Castin,Japan,Kawaguchi,Ketterle,Delgado}. In the present case, the difference is that the spatial modulation of the local nonlinearity in Eq. (\ref{gprofile}) induces an effective trap, which prevents the separation of the split pair, keeping it in the rotating state. The same trap maintains the stability of the double vortices in the blue area of Fig. \ref{fig7}(b) (the stabilization of multiple vortices by an external potential is known in other models \cite{stabilization}).

Lastly, the precession of vortices with $m=2$, which are initially placed off the center, is displayed in Figs. \ref{fig9}(d,e). In panels (d) and (e), the double vortices, with $\left( r_{0}=5,k=-5\right) $ and $\left(r_{0}=3,k=-2\right)$ are, respectively, initially unstable and stable against splitting, according to Fig. \ref{fig7}(b). In both cases, the shifted double vortices split in rotating pairs with a small separation between the unitary vortices, which perform robust precession. Further, the separation between the unitary vortices in the pair remains constant if the original double vortex is stable (Fig. \ref{fig9}(e)), while the size of the pair produced by splitting of an unstable double vortex slowly increases in
the course of the evolution, as seen in Figs. \ref{fig9}(c,d). The consideration of vortices with $m\geq 3$ is left beyond the scope of the present work.

\section{Conclusion}

\label{sec4}
We have considered the model of the optical medium or BEC with intrinsic self-repulsion. The model includes the spatial modulation of the local nonlinearity strength, with a maximum in the central area and a minimum in the periphery. The analysis has revealed new modes, \textit{viz}., the 1D and 2D flat-floor bubbles, 1D flat-waist dark solitons, and 2D flat-waist vortices. In a broad parameter region, the structure of the modes is accurately predicted by the TFA (Thomas-Fermi approximation). Through the computation of eigenvalues for small perturbations, it is predicted that the 1D and 2D flat-floor bubbles, as well as 2D vortices with winding number $m=1 $, are completely stable, which is corroborated by direct simulations. Nontrivial findings are instability boundaries, in their existence area, for dark solitons and vortices with $m=2$. Unstable dark solitons spontaneously escape into the peripheral area with a smaller local self-repulsion coefficient, which is explained by means of the analytical approximation. Unstable double vortices split in rotating pairs of unitary vortices. Lastly, originally displaced stable vortices feature robust precession around the origin.

The consideration of vortices with $m\geq 3$ may be interesting too, especially the precession in the presence of the flat-waist structure.

\section*{Funding}

The authors gratefully acknowledge financial support from the National Major Instruments and Equipment Development Project of National Natural Science Foundation of China (No. 61827815), the National Natural Science Foundation of China (No. 62075138), the Science and Technology Project of Shenzhen (Nos. JCYJ20190808121817100, JCYJ20190808164007485, JSGG20191231144201722),
and the Israel Science Foundation (grant No. 1286/17).

\section*{Compliance with ethical standards}

\section*{Conflict of interest}

The authors declare that they have no conflict of interest.

\section*{Data Availability Statements}

The datasets generated during and analysed during the current study are available from the corresponding author on reasonable request.


\begin{thebibliography}{0}%
\makeatletter
\providecommand \@ifxundefined [1]{%
 \@ifx{#1\undefined}
}%
\providecommand \@ifnum [1]{%
 \ifnum #1\expandafter \@firstoftwo
 \else \expandafter \@secondoftwo
 \fi
}%
\providecommand \@ifx [1]{%
 \ifx #1\expandafter \@firstoftwo
 \else \expandafter \@secondoftwo
 \fi
}%
\providecommand \natexlab [1]{#1}%
\providecommand \enquote  [1]{``#1''}%
\providecommand \bibnamefont  [1]{#1}%
\providecommand \bibfnamefont [1]{#1}%
\providecommand \citenamefont [1]{#1}%
\providecommand \href@noop [0]{\@secondoftwo}%
\providecommand \href [0]{\begingroup \@sanitize@url \@href}%
\providecommand \@href[1]{\@@startlink{#1}\@@href}%
\providecommand \@@href[1]{\endgroup#1\@@endlink}%
\providecommand \@sanitize@url [0]{\catcode `\\12\catcode `\$12\catcode
  `\&12\catcode `\#12\catcode `\^12\catcode `\_12\catcode `\%12\relax}%
\providecommand \@@startlink[1]{}%
\providecommand \@@endlink[0]{}%
\providecommand \url  [0]{\begingroup\@sanitize@url \@url }%
\providecommand \@url [1]{\endgroup\@href {#1}{\urlprefix }}%
\providecommand \urlprefix  [0]{URL }%
\providecommand \Eprint [0]{\href }%
\providecommand \doibase [0]{http://dx.doi.org/}%
\providecommand \selectlanguage [0]{\@gobble}%
\providecommand \bibinfo  [0]{\@secondoftwo}%
\providecommand \bibfield  [0]{\@secondoftwo}%
\providecommand \translation [1]{[#1]}%
\providecommand \BibitemOpen [0]{}%
\providecommand \bibitemStop [0]{}%
\providecommand \bibitemNoStop [0]{.\EOS\space}%
\providecommand \EOS [0]{\spacefactor3000\relax}%
\providecommand \BibitemShut  [1]{\csname bibitem#1\endcsname}%
\let\auto@bib@innerbib\@empty
%</preamble>
\end{thebibliography}%


\begin{thebibliography}{999}
\bibitem{review} Kartashov, Y.V., Malomed, B.A., Torner, L.: Solitons in nonlinear lattices. Rev. Mod. Phys. \textbf{83}, 247--305 (2011)

\bibitem{NLSE0} Malomed, B.A., Mihalache, D., Wise, F., Torner, L.: Spatiotemporal optical solitons. J. Opt. B \textbf{7}, R53--R72 (2005)

\bibitem{NLSEND} Wang, Q., Deng, Z.: Controllable propagation path of imaginary value off-axis vortex soliton in nonlocal nonlinear media. Nonlinear Dyn. \textbf{100}, 1589--1598 (2020)

\bibitem{MVRS} Zezyulin, D.A., Kartashov, Y.V., Skryabin, D.V., Shelykh, I.A.: Spin-orbit coupled polariton condensates in a radially periodic potential: Multiring vortices and rotating solitons. ACS Photonics \textbf{5}, 3634--3642 (2018)

\bibitem{NLSE3} Ivanov, S.K., Kartashov, Y.V., Szameit, A., Torner, L., Konotop, V.V.: Vector topological edge solitons in Floquet insulators. ACS Photonics \textbf{7}, 735--745 (2020)

\bibitem{RRP2017} Mihalache, D.: Multidimensional localized structures in optical and matter-wave media: A topical survey of recent literature. Rom. Rep. Phys. \textbf{69}, 403 (2017)

\bibitem{soliton-NRP} Kartashov, Y.V., Astrakharchik, G.E., Malomed, B.A., Torner, L.: Frontiers in multidimensional self-trapping of nonlinear fields and matter. Nat. Rev. Phys. \textbf{1}, 185--197 (2019)

\bibitem{RJP2019} Malomed, B.A., Mihalache, D.: Nonlinear waves in optical and matter-wave media: A topical survey of recent theoretical and experimental results. Rom. J. Phys. \textbf{64}, 106 (2019)

\bibitem{PTREV} Konotop, V.V., Yang, J., Zezyulin, D.A.: Nonlinear waves in $\mathcal{PT}$-symmetric systems. Rev. Mod. Phys. \textbf{88}, 035002 (2016)

\bibitem{NLSE1} Kartashov, Y.V., Konotop, V.V., Modugno, M., Sherman, E.Y.: Solitons in inhomogeneous gauge potentials: integrable and nonintegrable dynamics. Phys. Rev. Lett. \textbf{122}, 064101 (2019)

\bibitem{NLSE2} Kartashov, Y.V., Konotop, V.V.: Stable nonlinear modes sustained by gauge fields. Phys. Rev. Lett. \textbf{125}, 054101 (2020)

\bibitem{Morsch} Morsch, O., Oberthaler, M.: Dynamics of Bose-Einstein condensates in optical lattices. Rev. Mod. Phys. \textbf{78}, 179--215, (2006)

\bibitem{GAP0} Konotop, V.V., Salerno, M.: Modulational instability in Bose-Einstein condensates in optical lattices. Phys. Rev. A \textbf{65}, 021602(R) (2002)

\bibitem{Kiv1} Louis, P.J.Y., Ostrovskaya, E.A., Savage, C.M., Kivshar, Y.S.: Bose-Einstein condensates in optical lattices: Band-gap structure and solitons. Phys. Rev. A \textbf{67}, 013602 (2003)

\bibitem{GAP1} Ye. F., Mihalache, D., Hu, B., Panoiu, N.C.: Subwavelength plasmonic lattice solitons in arrays of metallic nanowires. Phys. Rev. Lett. \textbf{104}, 106802 (2010)

\bibitem{GAP2} Xie, X.T., Macovei, M.A.: Single-cycle gap soliton in a subwavelength structure. Phys. Rev. Lett. \textbf{104}, 073902 (2010)

\bibitem{GAP3} He, Y., Zhu, X., Mihalache, D., Liu, J., Chen, Z.: Lattice solitons in $\mathcal{PT}$-symmetric mixed linear-nonlinear optical lattices. Phys. Rev. A \textbf{85}, 013831 (2012)

\bibitem{GAPRRP} Wang, H., Ren, X., Mihalache, D., Weng, Y., Huang, D., He, Y.: Defect modes supported by parity-time-symmetric triangular optical lattices with self-defocusing Kerr nonlinearity. Rom. Rep. Phys. \textbf{71}, 411 (2019)

\bibitem{subfund} Mayteevarunyoo, T., Malomed, B.A.: Stability limits for gap solitons in a Bose--Einstein condensate trapped in a time-modulated optical lattice. Phys. Rev. A \textbf{74}, 033616 (2006)

\bibitem{GAP7} Desyatnikov, A.S., Ostrovskaya, E.A., Kivshar, Y.S., Denz, C.: Composite band-gap solitons in nonlinear optically induced lattices. Phys. Rev. Lett. \textbf{91}, 153902 (2003)

\bibitem{GAP5} Fan, Z., Chen, Z., Li, Y., Malomed, B.A.: Gap and embedded solitons in microwave-coupled binary condensates. Phys. Rev. A \textbf{101}, 013607 (2020)

\bibitem{GAP8} Kartashov, Y.V., Vysloukh, V.A., Torner, L.: Surface gap solitons. Phys. Rev. Lett. \textbf{96}, 073901 (2006)

\bibitem{HS1} Sakaguchi, H., Malomed, B.A.: Dynamics of positive- and negative-mass solitons in optical lattices and inverted traps. J. Phys. B \textbf{37}, 1443--1459 (2004)

\bibitem{GAP9} Islam, M.J., Atai, J.: Stability of moving gap solitons in linearly coupled Bragg gratings with cubic-quintic nonlinearity. Nonlinear Dyn. \textbf{91}, 2725--2733 (2018)

\bibitem{GAP10} Zeng, L., Zeng, J.: Gap-type dark localized modes in a Bose-Einstein condensate with optical lattices. Adv. Photon. \textbf{1}, 046004 (2019)

\bibitem{additional} Gorbach, A.V., Malomed, B.A., Skryabin, D.V.: Gap polariton solitons. Phys. Lett. A \textbf{373}, 3024--3027 (2009)

\bibitem{GAP11} Goblot, V., Rauer, B., Vicentini, F., Le Boit\'{e}, A., Galopin, E., Lema\^{\i}tre, A., Le Gratiet, L., Harouri, A., Sagnes, I., Ravets, S., Ciuti, C., Amo, A., Bloch, J.: Nonlinear polariton fluids in a flatband reveal discrete gap solitons. Phys. Rev. Lett. \textbf{123}, 113901 (2019)

\bibitem{GAP12} Zhu, X., Yang, F., Cao, S., Xie, J., He, Y.: Multipole gap solitons in fractional Schr\"{o}dinger equation with parity-time-symmetric optical lattices. Opt. Express \textbf{28}, 1631--1639 (2020)

\bibitem{GAP13} Zeng, L., Zeng, J.: Preventing critical collapse of higher-order solitons by tailoring unconventional optical diffraction and nonlinearities. Commun. Phys. \textbf{3}, 26 (2020)

\bibitem{GAP6} Wang, P., Zheng, Y., Chen, X., Huang, C., Kartashov, Y.V., Torner, L., Konotop, V.V., Ye, F.: Localization and delocalization of light in photonic moir\'{e} lattices. Nature \textbf{577}, 42--46 (2020)

\bibitem{GAPNP} Fu, Q., Wang, P., Huang, C., Kartashov, Y.V., Torner, L., Konotop, V.V., Ye, F.: Optical soliton formation controlled by angle twisting in photonic moir\'{e} lattices. Nat. Photon. \textbf{14}, 663--668 (2020)

\bibitem{GAP4} Zeng, L., Zeng, J.: One-dimensional gap solitons in quintic and cubic-quintic fractional nonlinear Schr\"{o}dinger equations with a periodically modulated linear potential. Nonlinear Dyn. \textbf{98}, 985--995 (2019)

\bibitem{Kiv2} Ostrovskaya, E.A., Kivshar, Y.S.: Matter-wave gap solitons in atomic band-gap structures. Phys. Rev. Lett. \textbf{90}, 160407 (2003)

\bibitem{HS2} Sakaguchi, H., Malomed, B.A.: Two-dimensional loosely and tightly bound solitons in optical lattices and inverted traps. J. Phys. B \textbf{37}, 2225--2239 (2004)

\bibitem{NL1} Konotop, V.V.: Small-amplitude envelope solitons in nonlinear lattices. Phys. Rev. E \textbf{53}, 2843--2858 (1996)

\bibitem{Abdullaev0} Abdullaev, F.K., Gammal, A., Tomio, L.: Dynamics of bright matter-wave solitons in a Bose-Einstein condensate with inhomogeneous scattering length. J. Phys. B \textbf{37}, 635--651 (2004)

\bibitem{Kevrekidis} Theocharis, G., Schmelcher, P., Kevrekidis, P.G., Frantzeskakis, D.J.: Matter-wave solitons of collisionally inhomogeneous condensates. Phys. Rev. A \textbf{72}, 033614 (2005)

\bibitem{Abdullaev1} Abdullaev, F.K., Garnier, J.: Propagation of matter-wave solitons in periodic and random nonlinear potentials. Phys. Rev. A \textbf{72}, 061605(R) (2005)

\bibitem{DENL} Zeng, L., Konotop, V.V., Lu, X., Cai, Y., Zhu, Q., Li, J.: Localized modes and dark solitons sustained by nonlinear defects. Opt. Lett. \textbf{46}, 2216--2219 (2021)

\bibitem{anti} Sakaguchi, H., Malomed, B.A.: Solitons in combined linear and nonlinear lattice potentials. Phys. Rev. A \textbf{81}, 013624 (2010)

\bibitem{NL2} Zeng, L., Zeng, J.: One-dimensional solitons in fractional Schr\"{o}dinger equation with a spatially periodical modulated nonlinearity: nonlinear lattice. Opt. Lett. \textbf{44}, 2661--2664 (2019)

\bibitem{CQNL} Zeng, L., Mihalache, D., Malomed, B.A., Lu, X., Cai, Y., Zhu, Q., Li, J.: Families of fundamental and multipole solitons in a cubic-quintic nonlinear lattice in fractional dimension. Chaos Solitons Fract. \textbf{144}, 110589 (2021)

\bibitem{NL3} Kartashov, Y.V., Malomed, B.A., Vysloukh, V.A., Torner, L.: Two-dimensional solitons in nonlinear lattices. Opt. Lett. \textbf{34}, 770--772 (2009)

\bibitem{EXP3} Wang, Z.Q., Nithyanandan, K., Coillet, A., Tchofo-Dinda, P., Grelu, Ph.: Optical soliton molecular complexes in a passively mode-locked fibre laser. Nat. Commun. \textbf{10}, 830 (2019)

\bibitem{EXP4} Weng, W., Bouchand, R., Lucas, E., Obrzud, E., Herr, T., Kippenberg, T.J.: Heteronuclear soliton molecules in optical microresonators. Nat. Commun. \textbf{11}, 2402 (2020)

\bibitem{EXP2} Liu, X., Yao, X., Cui, Y.: Real-time observation of the buildup of soliton molecules. Phys. Rev. Lett. \textbf{121}, 023905 (2018)

\bibitem{EXP1} Herink, G., Kurtz, F., Jalali, B., Solli, D.R., Ropers, C.: Real-time spectral interferometry probes the internal dynamics of femtosecond soliton molecules. Science \textbf{356}, 50--54 (2017)

\bibitem{EXP5} Kurtz, F., Ropers, C., Herink, G.: Resonant excitation and all-optical switching of femtosecond soliton molecules. Nat. Photon. \textbf{14}, 9--13 (2020)

\bibitem{SDN1} Borovkova, O.V., Kartashov, Y.V., Torner, L., Malomed, B.A.: Bright solitons from defocusing nonlinearities. Phys. Rev. E \textbf{84} 035602(R) (2011)

\bibitem{SDN3} Driben. R., Dror. N., Malomed, B.A., Meier, T.: Multipoles and vortex multiplets in multidimensional media with inhomogeneous defocusing nonlinearity. New J. Phys. \textbf{17}, 083043 (2015)

\bibitem{SDN2} Borovkova, O.V., Kartashov, Y.V., Malomed, B.A., Torner, L.: Algebraic bright and vortex solitons in defocusing media. Opt. Lett. \textbf{36}, 3088--3090 (2011)

\bibitem{SDN4} Lobanov, V.E., Borovkova, O.V., Kartashov, Y.V., Malomed, B.A., Torner, L.: Stable bright and vortex solitons in photonic crystal fibers with inhomogeneous defocusing nonlinearity. Opt. Lett. \textbf{37}, 1799--1801 (2012)

\bibitem{Jinhua} Zeng, J., Malomed, B.A.: Localized dark solitons and vortices in defocusing media with spatially inhomogeneous nonlinearity. Phys. Rev. E \textbf{95}, 052214 (2017)

\bibitem{SDN6} Driben, R., Meier, T., Malomed, B.A.: Creation of vortices by torque in multidimensional media with inhomogeneous defocusing nonlinearity. Sci. Rep. \textbf{5}, 9420 (2015)

\bibitem{SDN5} Driben, R., Kartashov, Y.V., Malomed, B.A., Meier, T., Torner, L.: Three-dimensional hybrid vortex solitons. New J. Phys. \textbf{16}, 063035 (2014)

\bibitem{Foshan} Zhong, R., Chen, Z., Huang, C., Luo, Z., Tan, H., Malomed, B.A., Li, Y.: Self-trapping under the two-dimensional spin-orbit-coupling and spatially growing repulsive nonlinearity. Front. Phys. \textbf{13}, 130311 (2018)

\bibitem{SDN7} Zeng, L., Zeng, J., Kartashov, Y.V., Malomed, B.A.: Purely Kerr nonlinear model admitting flat-top solitons. Opt. Lett. \textbf{44}, 1206--1209 (2019)

\bibitem{SDN8} Zeng, L., Zeng, J.: Gaussian-like and flat-top solitons of atoms with spatially modulated repulsive interactions. J. Opt. Soc. Am. B \textbf{36}, 2278--2284 (2019)

\bibitem{SDN9} Kartashov, Y.V., Malomed, B.A., Shnir, Y., Torner, L.: Twisted toroidal vortex solitons in inhomogeneous media with repulsive nonlinearity. Phys. Rev. Lett. \textbf{113}, 264101 (2014)

\bibitem{SDN10} Driben, R., Kartashov, Y.V., Malomed, B.A., Meier, T., Torner, L.: Soliton gyroscopes in media with spatially growing repulsive nonlinearity. Phys. Rev. Lett. 112, 020404 (2014)

\bibitem{SDN11} Kengne, E., Lakhssassi, A., Liu, W.: Non-autonomous solitons in inhomogeneous nonlinear media with distributed dispersion. Nonlinear Dyn. \textbf{97}, 449--469 (2019)

\bibitem{SDN12} Kartashov, Y.V., Malomed, B.A., Vysloukh, V.A., Beli\'{c}, M.R., Torner, L.: Rotating vortex clusters in media with inhomogeneous defocusing nonlinearity. Opt. Lett. \textbf{42}, 446--449 (2017)

\bibitem{SDN13} Zeng, L., Zeng, J.: Modulated solitons, soliton and vortex clusters in purely nonlinear defocusing media. Ann. Phys. \textbf{421}, 168284 (2020)

\bibitem{COLLAPSE} Berg\'{e}, L.: Soliton stability versus collapse. Phys. Rev. E 62, R3071--R3074 (2000)

\bibitem{Barash} Barashenkov, I.V., Panova, E.Y.: Stability and evolution of the quiescent and traveling solitonic bubbles. Physica D \textbf{69}, 114--134 (1993)

\bibitem{BUB1} Becker, C., Sengstock, K., Schmelcher, P., Kevrekidis, P.G., Carretero-Gonz\'{a}lez, R.: Inelastic collisions of solitary waves in anisotropic Bose-Einstein condensates: sling-shot events and expanding collision bubbles. New J. Phys. \textbf{15}, 113028 (2013)

\bibitem{BUB2} Varga, R., Pa\'{a}l, G.: Numerical investigation of the strength of collapse of a harmonically excited bubble. Chaos Solitons Fract. \textbf{76}, 56--71 (2015)

\bibitem{Cid} Falc\~{a}o-Filho, E.L., de Ara\'{u}jo, C.B., Boudebs, G., Leblond, H., Skarka, V.: Robust two-dimensional spatial solitons in liquid carbon disulfide. Phys. Rev. Lett. \textbf{110}, 013901 (2013)

\bibitem{Petrov1} Petrov, D.S.: Quantum mechanical stabilization of a collapsing Bose-Bose mixture. Phys. Rev. Lett. \textbf{115}, 155302 (2015)

\bibitem{Petrov2} Petrov, D.S., Astrakharchik, G.E.: Ultradilute low-dimensional liquids. Phys. Rev. Lett. \textbf{117}, 100401 (2016)

\bibitem{Leticia1} Cabrera, C.R., Tanzi, L., Sanz, J., Naylor, B., Thomas, P., Cheiney, P., Tarruell, L.: Quantum liquid droplets in a mixture of Bose-Einstein condensates. Science \textbf{359}, 301--304 (2018)

\bibitem{Inguscio} Semeghini, G., Ferioli, G., Masi, L., Mazzinghi, C., Wolswijk, L., Minardi, F., Modugno, M., Modugno, G., Inguscio, M., Fattori, M.: Self-bound quantum droplets in atomic mixtures in free space. Phys. Rev. Lett. \textbf{120}, 235301 (2018)

\bibitem{hetero} D'Errico, C., Burchianti, A., Prevedelli, M., Salasnich, L., Ancilotto, F., Modugno, M., Minardi, F., Fort, C.: Observation of quantum droplets in a heteronuclear bosonic mixture. Phys. Rev. Research \textbf{1}, 033155 (2019)

\bibitem{swirling} Kartashov, Y.V., Malomed, B.A., Tarruell, L., Torner, L.: Three-dimensional droplets of swirling superfluids. Phys. Rev. A \textbf{98}, 013612 (2018)

\bibitem{Raymond} Li, Y., Chen, Z., Luo, Z., Huang, C., Tan, H., Pang, W., Malomed, B.A.: Two-dimensional vortex quantum droplets. Phys. Rev. A \textbf{98}, 063602 (2018)

\bibitem{Denz} Woerdemann, M., Alpmann, C., Esseling, M., Denz, C.: Advanced optical trapping by complex beam shaping. Laser Photonics Rev. \textbf{7}, 839--854 (2013)

\bibitem{Bliokh} Bliokh, K.Y., Nori, F.: Transverse and longitudinal angular momenta of light. Phys. Rep. \textbf{592}, 1--38 (2015)

\bibitem{Gross} Gross, E.P.: Structure of a quantized vortex in boson systems. Nuovo Cim. \textbf{20}, 454--457 (1961)

\bibitem{Pit0} Pitaevskii, L.P.: Vortex lines in an imperfect Bose gas. Sov. Phys. JETP \textbf{13}, 451--454 (1961)

\bibitem{Pit} Pitaevskii, L.P., Stringari, S.: Bose-Einstein condensation. Oxford University Press, Oxford (2003)

\bibitem{Kivshar} Kivshar, Y.S., Luther-Davies, B.: Dark optical solitons: physics and applications. Phys. Rep. \textbf{298}, 81--197 (1998)

\bibitem{Dimitri} Frantzeskakis, D.J.: Dark solitons in atomic Bose--Einstein condensates: from theory to experiments. J. Phys. A \textbf{43}, 213001 (2010)

\bibitem{VK} Vakhitov, N.G., Kolokolov, A.A.: Stationary solutions of the wave equation in a medium with nonlinearity saturation. Radiophys. Quantum Electron. \textbf{16}, 783--789 (1973)

\bibitem{Berge} Berg\'{e} L.: Wave collapse in physics: principles and applications to light and plasma waves. Phys. Rep. \textbf{303}, 259--370 (1998)

\bibitem{Fibich} Fibich, G.: The Nonlinear Schr\"{o}dinger Equation: Singular Solutions and Optical Collapse. Springer, Heidelberg (2015)

\bibitem{Newton} Kelley, C.T.: Solving nonlinear equations with Newton's method. SIAM, Philadelphia (2003)

\bibitem{FDTD2} Antoine, X., Duboscq, R.: GPELab, a Matlab toolbox to solve Gross--Pitaevskii equations II: Dynamics and stochastic simulations. Comput. Phys. Commun. \textbf{193}, 95--117 (2015)

\bibitem{Champneys} Champneys, A.R.: Homoclinic orbits in reversible systems and their applications in mechanics, fluids and optics. Physica D \textbf{112}, 158-186 (1998)

\bibitem{pert} Kivshar, Y.S., Yang, X.P.: Perturbation-induced dynamics of dark solitons. Phys. Rev. E \textbf{49}, 1657--1670 (1994)

\bibitem{Busch} Busch, T., Anglin, J.R.: Motion of dark solitons in trapped Bose-Einstein condensates. Phys. Rev. Lett. \textbf{84}, 2298--2301 (2000)

\bibitem{precession} Anderson, B.P., Haljan, P.C., Wieman, C.E., Cornell, E.A.: Vortex precession in Bose-Einstein condensates: Observations with filled and empty cores. Phys. Rev. Lett. \textbf{85}, 2857--2860 (2000)

\bibitem{Skryabin} Skryabin, D.V.: Instabilities of vortices in a binary mixture of trapped Bose-Einstein condensates: Role of collective excitations with positive and negative energies. Phys. Rev. A \textbf{63}, 013602 (2000)

\bibitem{discr-vort} Malomed, B.A., Kevrekidis, P.G.: Discrete vortex solitons. Phys. Rev. E \textbf{64}, 026601 (2001)

\bibitem{Fetter} Fetter, A.L.: Rotating trapped Bose-Einstein condensates. Rev. Mod. Phys. \textbf{81}, 647--691 (2009)

\bibitem{Neu} Neu, J.C.: Vortices in complex scalar fields. Physica D \textbf{43}, 385--406 (1990)

\bibitem{Pu} Pu, H., Law, C.K., Eberly, J.H., Bigelow, N.P.: Coherent disintegration and stability of vortices in trapped Bose condensates. Phys. Rev. A \textbf{59}, 1533--1537 (1999)

\bibitem{Castin} Castin, Y., Dum, R.: Bose-Einstein condensates with vortices in rotating traps. Eur. Phys. J. D \textbf{7}, 399--412 (1999)

\bibitem{Japan} M\"{o}t\"{o}nen, M., Mizushima, T., Isoshima, T., Salomaa, M.M., Machida, K.: Splitting of a doubly quantized vortex through intertwining in Bose-Einstein condensates. Phys. Rev. A \textbf{68}, 023611 (2003)

\bibitem{Kawaguchi} Kawaguchi, Y., Ohmi, T.: Splitting instability of a multiply charged vortex in a Bose-Einstein condensate. Phys. Rev. A \textbf{70}, 043610 (2004)

\bibitem{Ketterle} Shin, Y., Saba, M., Vengalattore, M., Pasquini, T.A., Sanner, C., Leanhardt, A.E., Prentiss, M., Pritchard, D.E., Ketterle, W.: Dynamical Instability of a Doubly Quantized Vortex in a Bose-Einstein Condensate. Phys. Rev. Lett. \textbf{93} 160406 (2004)

\bibitem{Delgado} Mateo, A.M., Delgado, V.: Dynamical evolution of a doubly quantized vortex imprinted in a Bose-Einstein condensate. Phys. Rev. Lett. \textbf{97}, 180409 (2006)

\bibitem{stabilization} Huhtam\"{a}ki, J.A.M., M\"{o}tt\"{o}nen, M., Virtanen, S.M.M.: Dynamically stable multiply quantized vortices in dilute Bose-Einstein condensates. Phys. Rev. A \textbf{74}, 063619 (2006)

\end{thebibliography}
\end{document}